\documentstyle[prl,aps,twocolumn,floats]{revtex}
\begin{document}
\draft

\def\type{l}			
\def\figures{f}			
\def\ueberschriften{n}

\if\figures e
\input epsf
\fi
\if\figures f
\input epsf
\fi
\if\type t
\input ../definitions.tex 
\smallerformat  
\fi
\language=0

\def\xvec	{{\bf x}}
\def\yvec	{{\bf y}}
\def\zvec	{{\bf z}}
\def\alphavec	{{\bf\alpha}}
\def\rvec	{{\bf r}}
\def\Ddiff	{{\cal D}}
\def\Dtens	{{\bf D}}
\def\ddim	{{d}}
\def\g		{{g}}
\def\tstar	{{t^*}}
\def\rstar	{{r^*}}
\def\bvec	{{\bf b}}
\def\gvec	{{\bf g}}
\def\dt		{{\Delta t}}
\def\dr		{{\Delta r}}
\def\mod{\bmod}
\def\bino(#1,#2)  {\left({#1\atop #2}\right)}
\def\phivec	{{\bf \Phi}}
\def\sigmasq {{\sigma^2}}
\def\zcom	{{z}}
\def\Dcom	{{D}}
\def\Op1{{\cal L}}
\def\frac#1/#2{\leavemode\kern.1em\raise.5ex\hbox{%
\the\scriptfont0 #1}%
\kern-.1em/\kern-.15em\lower.25ex\hbox{\the\scriptfont0 #2}}

\title{A New Class of Cellular Automata for \\
		Reaction-Diffusion Systems}
\author{J\"org R. Weimar and Jean-Pierre Boon}
\address{Facult\'{e} des Sciences, C.P. 231\\
Universit\'{e} Libre de Bruxelles, B-1050 Bruxelles, Belgium\\
E-mail: {\tt jweimar@ulb.ac.be, jpboon@ulb.ac.be}
}
\date{May 22, 1993}
\maketitle

\begin{abstract}
We introduce a new class of cellular automata to model 
reaction-diffusion systems in a quantitatively correct way. 
The construction of the CA from the reaction-diffusion equation 
relies on a moving average procedure to implement diffusion, 
and a probabilistic table-lookup for the reactive part. 
The applicability of the new CA is demonstrated using the 
Ginzburg-Landau equation.
\end{abstract}

\narrowtext

\if\ueberschriften j 
\section{Introduction}
\label{sec:intro}
\fi
Cellular automata models have been used in many applications to 
model reactive and diffusive systems 
\cite{Wolfram86:0,Kapral91:113}.
Most uses of cellular automata (CAs) can be classified into one of 
four approaches:
(i) Ising-type models of phase transitions; 
(ii) lattice gas models (the lattice gas method was initially 
developed to 
model hydrodynamic flows and has been extended in many directions 
\cite{Boon92:0,DoolenBook});
(iii) systematic investigation of the behavior of CAs by 
investigating 
{\it all} rules of a certain class (e.g., all possible rules 
for one-dimensional automata with two states and 
nearest neighbor interaction)
\cite{Wolfram86:0};
and (iv) qualitative discrete modelling 
(including operational use of CAs as 
an alternative to partial differential equations (PDEs) 
\cite{Toffoli84:117}). 
These models are generally based on qualitative rather than 
quantitative information about the system to be modelled.
A CA is constructed which preserves the qualitative features deemed 
most relevant and it is then investigated 
(The lattice gas methods were 
also developed along these lines \cite{Lawniczak91:132}).
Existing CA models for reaction-diffusion systems 
\cite{Greenberg78:515,Weimar92:309,Weimar92:328,Markus90:56,Schepers92:337} 
fall into the category (iv), i.e., they show 
qualitatively ``correct'' behavior and are restricted to certain 
reaction-diffusion (R-D) models and certain types of phenomena. 
This is the main criticism 
of experimentalists and researchers working with partial 
differential equation models, 
who search for quantitative predictions. 
In this letter we describe a class of CAs which is suitable for 
modelling many reaction-diffusion systems in a quantitatively 
correct way.
The new CAs are operationally more efficient than the reactive 
lattice gas methods, which also achieve quantitative correctness.
\if\ueberschriften j 
In section \ref{sec:automaton} we describe the construction of 
the new class of CAs and in section \ref{sec:ginzburg} we 
present the automaton using the Ginzburg-Landau equation as 
an example.
\fi
\if\ueberschriften n 
We first describe the construction of the new class of CAs; 
then we present the automaton using the Ginzburg-Landau 
equation as an example.
\fi

\if\ueberschriften j 
\section{A New Cellular Automaton Class}
\label{sec:automaton}
\fi
The main idea behind this class of CAs is careful discretization. 
Space and time are discretized as in normal finite difference methods 
for solving the PDE's. Finite difference methods then proceed to 
solve the resulting coupled system of $N \times s$ ordinary 
differential equations ($N$ points in space, 
$s$ equations in the PDE system) by any of a number of numerical 
methods, operating on floating point numbers. The use of floating 
point numbers on computers implies a discretization of the 
continuous variables. The errors introduced by this discretization 
and the ensuing roundoff errors are often not considered explicitly, 
but assumed to be small because the precision is rather high 
(8 decimal digits for usual floating point numbers). 
In contrast, in the CA approach, all variables are explicitly 
discretized into relatively small integers. This discretization 
allows the use of lookup tables to replace the evaluation of the 
nonlinear rate functions. It is this table lookup, combined with 
the fact that all calculations are performed using integers instead of 
floating point variables, that accounts for an improvement in 
speed of orders of magnitude on a conventional multi-purpose computer. 
The undesirable effects of discretization are overcome by using 
probabilistic rules for the updating of the CA.

The state of the CA is given by a regular array of concentration 
vectors $\yvec$ residing on a $d$-dimensional lattice. 
Each $\yvec(\rvec)$ is a $s$-vector of integers 
($s$ is the number of reactive species). For reasons of efficiency, and to 
fulfill the finiteness condition of the definition of cellular automata, 
each component $\yvec_i(\rvec)$ can only take integer values 
between 0 and $\bvec_i$, where the $\bvec_i$'s can be different 
for each species $i$. The position index $\rvec$ is a 
$\ddim$-dimensional vector in the CA lattice. For cubic lattices, 
$\rvec$ is a $\ddim$-vector of integers. 


The central operation of the automaton consists of calculating the sum
\begin{equation}
\tilde \yvec_i(\rvec) = \sum_{\rvec'\in N_i} \yvec_i(\rvec + \rvec')
\label{classcabeqa}
\end{equation}
of the concentrations in some neighborhood $N_i$. The neighborhoods 
can be different for each species $i$. A neighborhood is specified 
as a set of displacement vectors, e.g. in two dimensions
\begin{eqnarray}
&N_{\rm 5-star} = \left\{ \left( 0 \atop 0\right),
	\left( 1 \atop 0\right),\left( 0 \atop 1\right),
	\left( -1 \atop 0\right),\left( 0 \atop -1\right)
\right\} \\
\noalign{\hbox{or}}
\lefteqn{N_{\rm 1-square} = }\nonumber\\
&N_{\rm 5-star} \cup \left\{ 
	\left( 1 \atop 1\right),\left( 1 \atop -1\right),
	\left( -1 \atop 1\right),\left( -1 \atop -1\right)
\right\}. \label{classcabeqb} 
\end{eqnarray}
For diffusive systems, symmetry requires that 
\begin{equation}	
\rvec' \in N \Rightarrow -\rvec' \in N
\end{equation}
and if, as is usually the case in reaction-diffusion systems, 
isotropy is required, then
\begin{equation}
	(r_1,\cdots,r_d)^T \in N \Rightarrow (r_{\alpha_1},
	\cdots,r_{\alpha_d})^T \in N
\end{equation}
for all index permutations $(\alpha_1,\cdots,\alpha_d)$.
For some neighborhoods, the summing operation can be executed 
in an extremely efficient way by using moving averages: 
in one dimension, the sum of all $2k+1$ cells centered around cell $r$
can be computed from the corresponding sum centered around 
cell $r-1$ with just one addition and subtraction:
\begin{eqnarray}
\lefteqn{\tilde\yvec_i(r) =\sum_{|r'| \le k} \yvec_i(r + r')} \nonumber\\
&=& \yvec_i(r-k) + \yvec_i(r-k+1)+ \cdots+ \yvec_i(r+k) \nonumber\\
&=& -\yvec_i(r-k-1) + \yvec_i(r-k-1) + \yvec_i(r-k)+\nonumber\\
&& \cdots+\yvec_i(r+k-1)+ \yvec_i(r+k) \nonumber\\
&=& -\yvec_i(r-k-1) + \tilde \yvec_i(r-1) + \yvec_i(r+k).
\label{classcabeqc}
\end{eqnarray}
Using this relationship recursively, and using the fact that the 
sum over a square neighborhood in $d$ dimensions can be constructed 
as the convolution of such one-dimensional sums applied in each 
dimension in turn, one obtains an algorithm to compute the sum of 
all cells in a square (cubic) neighborhood with only $2 d$ 
additions per cell \cite{Weimar92:309}.
In a multispecies model several variables are necessary, but they 
can be packed into one 
computer word. In this case the averaging operation can be performed 
on several species at once if the diffusion coefficients are equal.

In the following we use the normalized values 
$\xvec_i(\rvec) = \yvec_i(\rvec)/\bvec_i$ 
and 
$\tilde \xvec_i(\rvec) = \tilde \yvec_i(\rvec)/(\bvec_i |N_i|)$,
which are always between zero and one. 
The resulting fields $\tilde \xvec_i(\rvec)$ are then the local averages 
of the $\xvec_i(\rvec)$. {\it The averaging has the effect of diffusion.}
This can be seen from a Taylor expansion of $\xvec_i(\rvec+\rvec')$ 
around $\xvec_i(\rvec)$:
\begin{eqnarray}
	\tilde \xvec_i(\rvec) &=& {1\over |N_i|} \sum_{\rvec'\in N} 
	\sum_{\kappa=0}^{\infty}
	{1\over \kappa!}\left(\rvec' 
		{\partial \over \partial \rvec'} \right)^\kappa 
	\xvec_i(\rvec) \nonumber\\
		&=& \xvec_i(\rvec) + D_i \nabla^2 \xvec_i(\rvec) + \cdots
\label{classcabeqf}
\end{eqnarray}
The factors $D_i$ can be computed as shown in \cite{Weimar92:309} and are
easily calculated from (\ref{classcabeqf}) for square 
neighborhoods with radius $k$: $D_i = k (k+1)/6$.

The second operation in the cellular automaton is the implementation of the 
reactive processes described by a rate law. 
Given the reaction-diffusion equation
\begin{equation}
{\partial \zvec \over \partial t} = \phivec(\zvec) + \Ddiff \nabla^2 \zvec
\end{equation}
(where $\nabla^2 $ is understood to be a spatial operator acting on 
each component of the vector $\zvec$ separately, and $\Ddiff$ is a 
diagonal matrix), we discretize the time derivative to obtain
\begin{equation}
\zvec^{\tstar+\dt} = \zvec^{\tstar} + \dt \phivec(\zvec^{\tstar})+ 
	\dt \Ddiff \nabla^2 \zvec^{\tstar} .
\end{equation}
Changing the time and space scales by setting $t = \tstar / \dt$ and  
$r = \rstar / \dr$, and using the variable $\xvec$ for the rescaled set gives 
\begin{equation}
\xvec^{t+1} = \xvec^{t} + \dt 
	\phivec(\xvec^{t})+ {\dt\over \dr^2} \Ddiff \nabla^2 \xvec^{t}.
\label{classcabeqd}
\end{equation}
as the equation to be treated by the CA. Let us define
\begin{equation}
\phivec^*(\xvec) = \xvec + \dt \phivec(\xvec).
\label{classcabeqe}
\end{equation}
From eqns. (\ref{classcabeqf}) and (\ref{classcabeqe})
\begin{eqnarray}
\phivec^*(\tilde\xvec^t) &=& \xvec^{t} + D_i \nabla^2 \xvec^t + \cdots +
	\dt \phivec(\xvec^{t} + D_i \nabla^2 \xvec^t + \cdots) \nonumber\\
	&=& \xvec^{t} + D_i \nabla^2 \xvec^t +
	\dt \phivec(\xvec^{t}) + {\rm O}(\dt^2).
\end{eqnarray}
Then
\begin{equation}
\xvec^{t+1} = \phivec^*(\tilde\xvec^t)
\label{classcaupdateing}
\end{equation}
is consistently first order accurate in time and within this 
limit, eq. (\ref{classcaupdateing}) can be validly identified 
with eq. (\ref{classcabeqd})  to describe the evolution of the 
system. The identification yields 
\begin{equation}
D_i = {\dt\over \dr^2} \Ddiff_{ii} \hbox{\qquad( $i=1,\cdots,s$)}
\end{equation}
which defines the space scale.
As $\tilde \yvec$ is the result of the diffusion step, the average 
output of the CA reaction-diffusion process should therefore be given by
\begin{equation}
	\gvec_j = \bvec_j \phivec^*\left(\left\{ 
	{\tilde\yvec_i\over \bvec_i |N_i|} \right\}_{i=1}^s\right)
\end{equation}
for species $j$.
\footnote{Possibly the function $\phivec^*$ needs to be 
truncated to conform to the condition $\phivec^* \in [0,1]^s$.}

A simplistic discretization of the rate law may produce 
problems: due to the discretization spurious steady
states or  oscillations can appear.  It is at this stage
that the probabilistic rules come  into effect. Given an
input configuration $\tilde \yvec^t(\rvec)$, one 
assigns new values $\yvec^{t+1}(\rvec)$ 
probabilistically in such a  way that the {\it average}
result corresponds to the finite difference
approximation to the given reaction-diffusion
equation, $\gvec_i$. The simplest CA rule for the
reactive step is to treat each species  separately and use 
$\yvec_j^{t+1} = \left\lfloor
\gvec_j\right\rfloor+1 $  with probability
$\left(\gvec_j\mod 1\right)$  and $\yvec_j^{t+1} =
\left\lfloor \gvec_j\right\rfloor$   otherwise
($\lfloor e \rfloor$ is the largest integer  smaller than
or equal to $e$). In this way the average is exactly 
$\gvec_j$. Clearly this method introduces the minimal 
amount of noise for a given set of $\bvec_i$'s. In case
higher noise  levels are desired (e.g., if one wants to
evaluate the role of  fluctuations
\cite{Weimar92:627}), one can choose a different  set of
output values  and associated probabilities with the
same average  $\gvec_j$, but different variance. To do
so, in general at least three different possible
outcomes (e.g.$\left\lfloor
\gvec_j\right\rfloor-1,\left\lfloor
\gvec_j\right\rfloor, \left\lfloor
\gvec_j\right\rfloor+1$) are necessary. 

\if\ueberschriften j 
\section{The new CA applied to the Ginzburg-Landau Equation}
\label{sec:ginzburg}
\fi
To demonstrate the applicability of the new CA we show how the 
Ginzburg-Landau equation can be mapped onto the automaton. 
We consider the PDE\cite{Kuramoto84:0,Hagan82:762}
\begin{equation}
{\partial \zcom \over \partial t} = 
\Dcom \nabla^2 \zcom + a \zcom - b |\zcom|^2 \zcom,
\end{equation}
which can be viewed (for $\Dcom$ real) as a two-species 
reaction-diffusion system by separating real and imaginary parts as 
$\zcom = x + i y $, $a = \alpha + i \gamma$, and $b = \beta + i \delta$,
\begin{mathletters}
\label{cglPDE}
\begin{eqnarray}
{\partial x \over \partial t} &=& D \nabla^2 x + 
	\alpha x - \gamma y + (- \beta x  + \delta y) (x^2+y^2)\\
{\partial y \over \partial t} &=& D \nabla^2 y + 
	\alpha y + \gamma x +( - \beta y  - \delta x) (x^2+y^2)
\end{eqnarray}
\end{mathletters}
where $x$, $y$ are space- and time-dependent variables. 
With $\zcom=r e^{i\phi}$, the corresponding rate equation can be 
conveniently written as
\begin{mathletters}
\begin{eqnarray}
{\partial r \over \partial t} &=& \alpha r - \beta r^3 = 
		\beta r \left(\alpha/\beta - r^2\right) \\
{\partial \phi \over \partial t} &=& \gamma - \delta r^2.
\end{eqnarray}
\end{mathletters}
One can set $\beta=1$ by changing the time scale. 
The stable homogeneous solution is $\zcom = r_s e^{i \Omega t}$ 
with $r_s = \sqrt{\alpha/\beta}$ and $\Omega = \gamma - \delta r_s^2$
(A steady but unstable solution is $\zcom = 0$).
Since the equation for $r$ is independent of $\phi$, one can 
transform the solution for $\Omega=0$ to any given $\Omega$ by 
multiplying $\zcom(t)$ with a factor $e^{i\omega t}$, yielding 
an oscillating solution.
The full PDE system, eq. (\ref{cglPDE}), also admits oscillating 
and rotating spirals, and other inhomogeneous solutions.

For the CA simulations, the region of interest in
concentration  space needs to be included in the unit
square $[0,1]^2$ by shifting  the origin to $(0.5,0.5)$
and scaling the parameters in such a way that  $ 0 < R_s < 0.5 $
(Here $R_s = 0.25$ and $\beta =1$). In all CA simulations
we use a square neighborhood (radius $k=1$), i.e.,  $D_1 =
D_2 = 1/3$, and discretization levels $\bvec_1 = \bvec_2
= 55$ (which gives a lookup-table size of 9 Mbytes).

Study of the homogeneous solutions shows that the CA
behaves like an explicit finite difference method with
added noise. The time step must be sufficiently small to
reproduce the correct solution. The noise, which is
intrinsic in the automaton and arises from the
discretization, has little effect on the amplitude of
the solution, but it introduces random drifts in the
phase.

We now turn to more interesting cases by considering
spiral wave solutions. Without loss of generality we use
$\Omega=0$, which eliminates the homogeneous
oscillations and thereby allows for larger time steps in
the computation and better visualization. 

\if\figures f
\begin{figure}
\epsfxsize=7.5truecm
\centerline{\epsffile{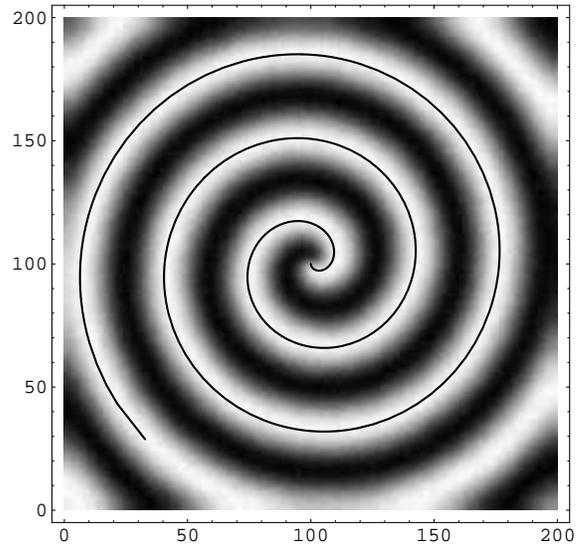}}
\caption{Spiral wave for $\Delta t = 2$, system size $200^2$ cells.
Superimposed on the grayscale plot of $x$ is the 
fitted Archimedian spiral.}
\label{fig:spiral}
\end{figure}
\fi

We initiate a spiral in a two-dimensional simulation by starting 
with the initial condition 
\begin{mathletters}
\begin{eqnarray}
R(r_x,r_y,t^0)  & =& c_1 \sqrt{(r_x-r_x^0)^2+(r_y-r_y^0)^2} \\
\theta(r_x,r_y,t^0)  & =& c_2 \arctan{r_y-r_y^0\over r_x-r_x^0} ,
\end{eqnarray}
\end{mathletters}
which creates exactly one phase singularity at
$(r_x^0,r_y^0)$. For nonzero $\delta$ smaller than a
critical value, a spiral develops and rotates steadily
after some time. In Figure \ref{fig:spiral} we show the
spiral obtained for  $\delta=1$, $\Delta t = 2$, and
system size $(200\Delta x)^2$. Notice that in spite of
the use of a square neighborhood for the CA, the spiral is
perfectly round. While this feature is expected when the
CA is viewed as a numerical method for solving the PDE, it
is not universally accepted that CA dynamics can produce
such an isotropic behavior.
 
\if\figures f
\begin{figure}
\epsfxsize=7truecm
\epsffile{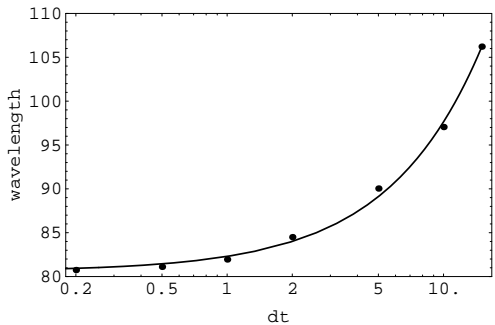}
\caption{Spiral wavelength as a function of the size of the 
time step $\Delta t$ (logarithmic scale). 
The data can be reasonably approximated 
as $\lambda \approx 80.6 + 1.7 \Delta t$ (solid curve).
The size of the system is selected to be 
$(6 \lambda_0)^2 \approx 490^2$ in non-dimensional units, 
corresponding to $490/ \protect\sqrt{3 \Delta t}$ cells
and the spiral fit is performed at t=4500.}
\label{fig:lamvsdt}
\end{figure}
\fi

In order to determine the effect of using large time
steps, we measure the asymptotic wavelength $\lambda$
by fitting an Archimedian spiral to the contour
$x<0,y=0$. The measured wavelengths for different time
steps $\Delta t$ are shown in Figure \ref{fig:lamvsdt}.
The value expected from the theory described in
\cite{Hagan82:762} is around $\lambda^0=81$. We
observe that the expected value is reached for small
enough $\Delta t$. Even for bigger $\Delta t$ the
deviations are relatively small and approximately
linear in $\Delta t$.  When the parameter $\delta$
becomes bigger than some critical value (estimated to be
$\delta_c=\beta\ 1.397\dots$ in
\cite{Hagan82:762}), the spiral wave solution becomes
unstable.  

As an example of greater complexity, Figure
\ref{fig:manyspirals} shows the simulation of a large
system ($500^2$cells) initialized in the state
$\zcom\approx 0$:  Under such conditions many
interacting spirals develop. Indeed, the initial state
is unstable, and because of the intrinsic (low level)
noise, different regions depart from this unstable
state with different phase values $\phi$. This
situation automatically creates many phase
singularities (points with $r=0$, surrounded by points
with all values of $\phi$), which then develop into
spirals. These phase singularities can merge and move as
they are influenced by each other\cite{Wu91:421}. 

\if\figures f
\begin{figure}
\epsfxsize=8.6truecm
\centerline{\epsffile{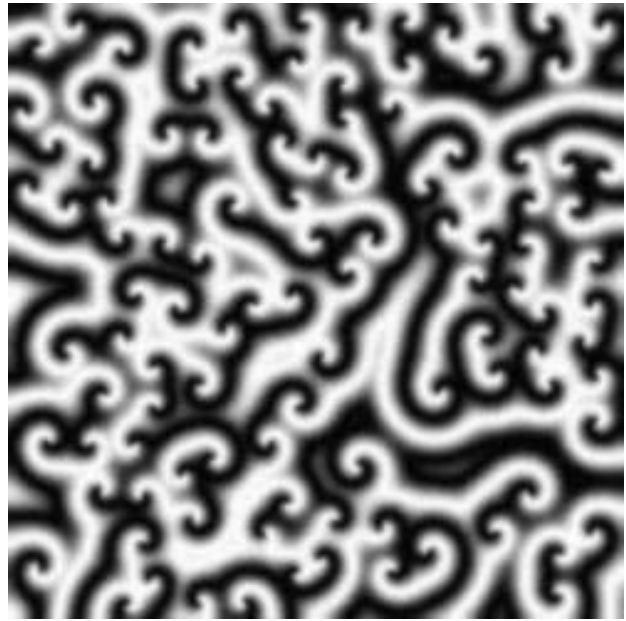}}
\caption{Collection of many interacting spirals in a large domain 
($500^2$cells, $\delta=1$, $\Delta t=2$, t=20000) }
\label{fig:manyspirals}
\end{figure}
\fi

Such large simulations are made possible by the speed
advantage that the CA offers over other numerical
methods for solving PDEs. We found that on a NeXT
workstation the CA runs about 5 times faster than an Euler
integration using the same time step. This advantage is
due to the use of integer arithmetic, table lookup, and
the fact that the diffusion calculation is faster in the
CA than in finite difference methods. The speed
advantage is much more pronounced when the nonlinear
rate functions are more complicated, as the table lookup
operation does not slow down.

In conclusion, we have constructed a class of cellular
automata that can be used to model many
reaction-diffusion systems with quantitatively
correct results. The solutions are very isotropic
despite the discreteness and the anisotropic nature of
the CA. The adverse effects of discretization are
overcome by the use of probabilistic rules. We have
demonstrated the applicability of the automaton for the
Ginzburg-Landau equation. We have also successfully
applied the method to more than ten different
reaction-diffusion models, some of which will be
reported in a forthcoming paper.

We acknowledge support from the European Community
(SC1-0212), Fonds National de la Recherche
Scientifique (JPB), Gottlieb Daimler- und Karl
Benz-Stiftung (JRW) and Stiftung Stipendien-Fonds des
Verbandes der Chemischen Industrie (JRW). 



\if\figures f
\else
\begin{figure}
\if\figures e
\epsfxsize=8.6truecm
\epsffile{spiral1.eps}
\fi
\caption{Spiral wave for $\Delta t = 2$, system size $200^2$ cells.
Superimposed on the grayscale plot of $x$ is the 
fitted Archimedian spiral.}
\label{fig:spiral}
\end{figure}

\begin{figure}
\if\figures e
\epsfxsize=8.6truecm
\epsffile{figlamvsdt.eps}
\fi
\caption{Spiral wavelength as a function of the size of the 
time step $\Delta t$ (logarithmic scale). 
The data can be reasonably approximated 
as $\lambda \approx 80.6 + 1.7 \Delta t$ (solid curve).
The size of the system is selected to be 
$(6 \lambda_0)^2 \approx 490^2$ in non-dimensional units, 
corresponding to $490/ \protect\sqrt{3 \Delta t}$ cells
and the spiral fit is performed at t=4500.}
\label{fig:lamvsdt}
\end{figure}

\begin{figure}
\if\figures e
\epsfxsize=8.6truecm
\epsffile{spirals2.eps}
\fi
\caption{Collection of many interacting spirals in a large domain 
($500^2$cells, $\delta=1$, $\Delta t=2$, t=20000) }
\label{fig:manyspirals}
\end{figure}
\fi

\end{document}